# Thermal Stability of Skyrmion Tubes in Nanostructured Cuboids

*Jialiang Jiang[1,2#], Jin Tang[1,2#]*, Tian Bai[3], Yaodong Wu[4], Jiazhuan Qin[3], Weixing Xia[3]*, Renjie Chen[3], Aru Yan[3], Shouguo Wang[5], Mingliang Tian[1,2], and Haifeng Du[2]**

[1]School of Physics and Optoelectronic Engineering, Anhui University, Hefei, 230601, China

[2]Anhui Province Key Laboratory of Condensed Matter Physics at Extreme Conditions, High Magnetic Field Laboratory, HFIPS, Anhui, Chinese Academy of Sciences, Hefei, 230031, China

[3]CISRI & NIMTE Joint Innovation Center for Rare Earth Permanent Magnets, Ningbo Institute of Material Technology and Engineering, Chinese Academy of Science, Ningbo 315201, China

[4]School of Physics and Materials Engineering, Hefei Normal University, Hefei, 230601, China

[5]Anhui Key Laboratory of Magnetic Functional Materials and Devices, School of Materials Science and Engineering, Anhui University, Hefei 230601, China

*Corresponding author: jintang@ahu.edu.cn; xiawxing@nimte.ac.cn; duhf@hmfl.ac.cn

[#]These authors contributed equally.

**Abstract**

Magnetic skyrmions in bulk materials are typically regarded as two-dimensional structures. However, they also exhibit three-dimensional configurations, known as skyrmion tubes, which elongate and extend in-depth. Understanding the configurations and stabilization mechanism of skyrmion tubes is crucial for the development of advanced spintronic devices. However, the generation and annihilation of skyrmion tubes in confined geometries are still rarely reported. Here, we present direct imaging of skyrmion tubes in nanostructured cuboids of a chiral magnet FeGe using Lorentz transmission electronic microscopy (TEM), while applying an in-plane magnetic field. It is observed that skyrmion tubes stabilize in a narrow field-temperature region near the Curie temperature ($T_c$). Through a field cooling process, metastable skyrmion tubes can exist in a larger region of the field-temperature diagram. Combining these experimental findings with micromagnetic simulations, we attribute these phenomena to energy differences and thermal fluctuations. Our results could promote topological spintronic devices based on skyrmion tubes.

Magnetic skyrmions are localized topological spin swirls that are potentially applicable in developing spintronic devices, due to their intriguing electromagnetic properties.[1-12] To achieve high-density skyrmion-based spintronic devices, it is necessary to investigate the geometrical confinement effect of nanostructures related to these devices.[3, 6, 12, 13] Previous studies have demonstrated the stabilization of skyrmions in nanodisks,[14, 15] nanostripes,[16] triangles,[17] and tetrahedrons.[18] Skyrmions exhibit remarkable stability and emergent topological magnetism in confined low-dimensional geometries.[14-21]

Although skyrmions have traditionally been considered two-dimensional magnetic objects,[22-25] recent studies have unveiled the existence of three-dimensional extended skyrmions with diverse morphologies and topological magnetism,[26-38] opening up new possibilities for the advancement of topological spintronic devices. The three-dimensional skyrmionic textures can be directly visualized using three-dimensional tomography[34, 35] or applying in-plane magnetic fields to thin lamella.[29, 31, 33, 36, 37] Understanding the fundamental physical principles behind the stabilization of skyrmion tubes, also referred to as skyrmion strings,[31, 33, 39-44] in strongly confined nanostructures is crucial for the exploration of small and functional topological spintronic devices, *e.g.*, the resistance abnormality of skyrmion tubes in nanowires[45] and the potential use of skyrmion tubes as nonplanar magnonic waveguides in nanostructured cuboids.[46] However, direct real-space imaging for the formation and stabilization of skyrmion tubes in confined nanostructured cuboids has not been explored yet.

Here, we present our findings on the stabilization of skyrmion tubes in nanostructured cuboids of a $B_{20}$ chiral FeGe magnet. We achieved this by applying in-plane magnetic fields using Lorentz transmission electronic microscopy. Our results reveal the confinement effects on the stabilization of skyrmion tubes and the thermal-field diagram of spin configurations in nanostructured cuboids. These findings could facilitate the development of tube-based topological spintronic device applications.

Magnetic configurations in FeGe cuboids are explored using Lorentz-TEM, which images the integral in-plane magnetic magnetizations along the depth orientation. However, it can only visualize the in-plane magnetization perpendicular to the tube direction, as the out-of-plane magnetic fields are typically generated by the object lens of TEM. Here, we develop an *in-situ* in-plane magnetic field generator, as shown in Figure 1. The in-plane magnetic field along the *x*-axis with a maximum magnitude of 118 mT is provided by a homemade electromagnet when passing through a current (see Figure 1a and supplemental note). The width of the in-plane field region, which is the gap between the two poles of the electromagnet, is approximately 30 μm. This gap allows for the insertion of nano-micro structures (Figure 1b).

The cubic-structured FeGe with $P2_13$ space group exhibits a high $T_c$ of approximately 278 K (Figure S1c).[47-49] At zero magnetic field, the ground state of FeGe is a spin spiral with a period $L_D \approx 70$ nm (Figure S1b).[21, 48, 50] Due to its weak magnetic anisotropy,[21, 51, 52] the $q$-vector does not show a preferred orientation (Figure S1b) in a large lamella.[53] To investigate the magnetic evolutions of skyrmion tubes in the confined geometry of FeGe, we fabricated a series of nanostructured (101) cuboids with different lengths $l$ using a focused ion beam (FIB) (Figure S1a). The lengths included 300, 510, 700, and 830 nm, while all of these cuboids had a similar cross-sectional size of approximately 220 nm × 160 nm ($w \times d$) (Figure 1c). This sample was placed in the gap of the electromagnet of the *in-situ* sample holder (Figure 1a, b). The magnetic evolutions of these cuboids were investigated in detail under the in-plane magnetic field. N. Mathur *et al.* discuss the magnetic transformations observed in a large thin FeCoGe lamella under an in-plane field of about 20 mT.[33] Here, using a homemade holder, we apply a larger in-plane field (up to 118 mT) to nanostructured cuboids, which enables the exploration of geometrical confinement effects and skyrmion tube stabilization in a broader field-temperature range.

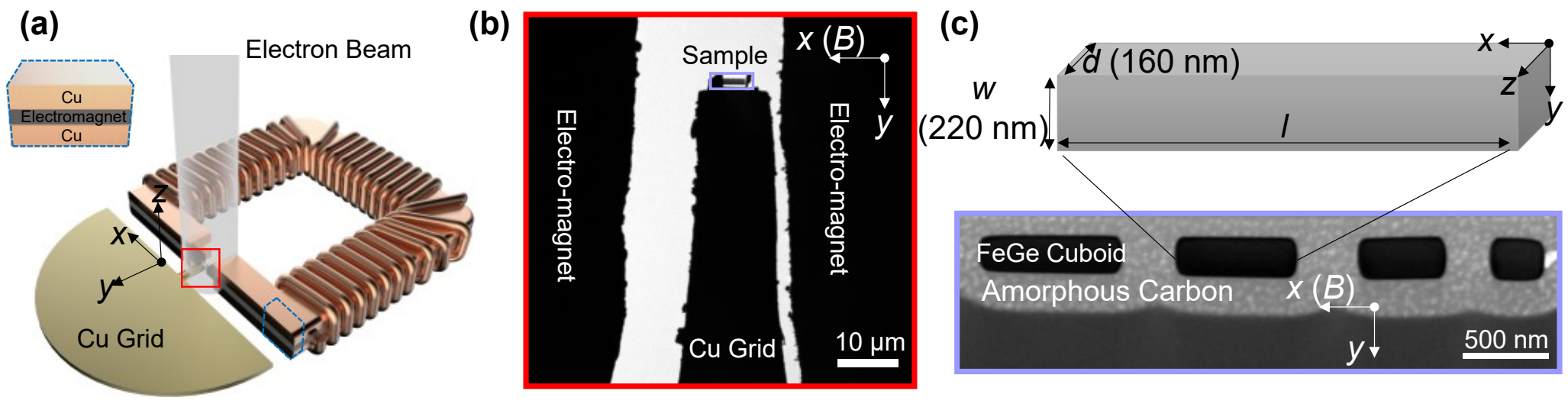

**Figure 1.** Nanostructured FeGe cuboids in an electromagnet of the in-situ magnetizing sample holder. (a) Schematic diagram of the in-situ device. Inset: The three-layer structure of the field generator. (b) TEM image of the structures marked by the red rectangle in (a). (c)

TEM image of the nanostructured FeGe cuboids, marked by the violet rectangle in (b).

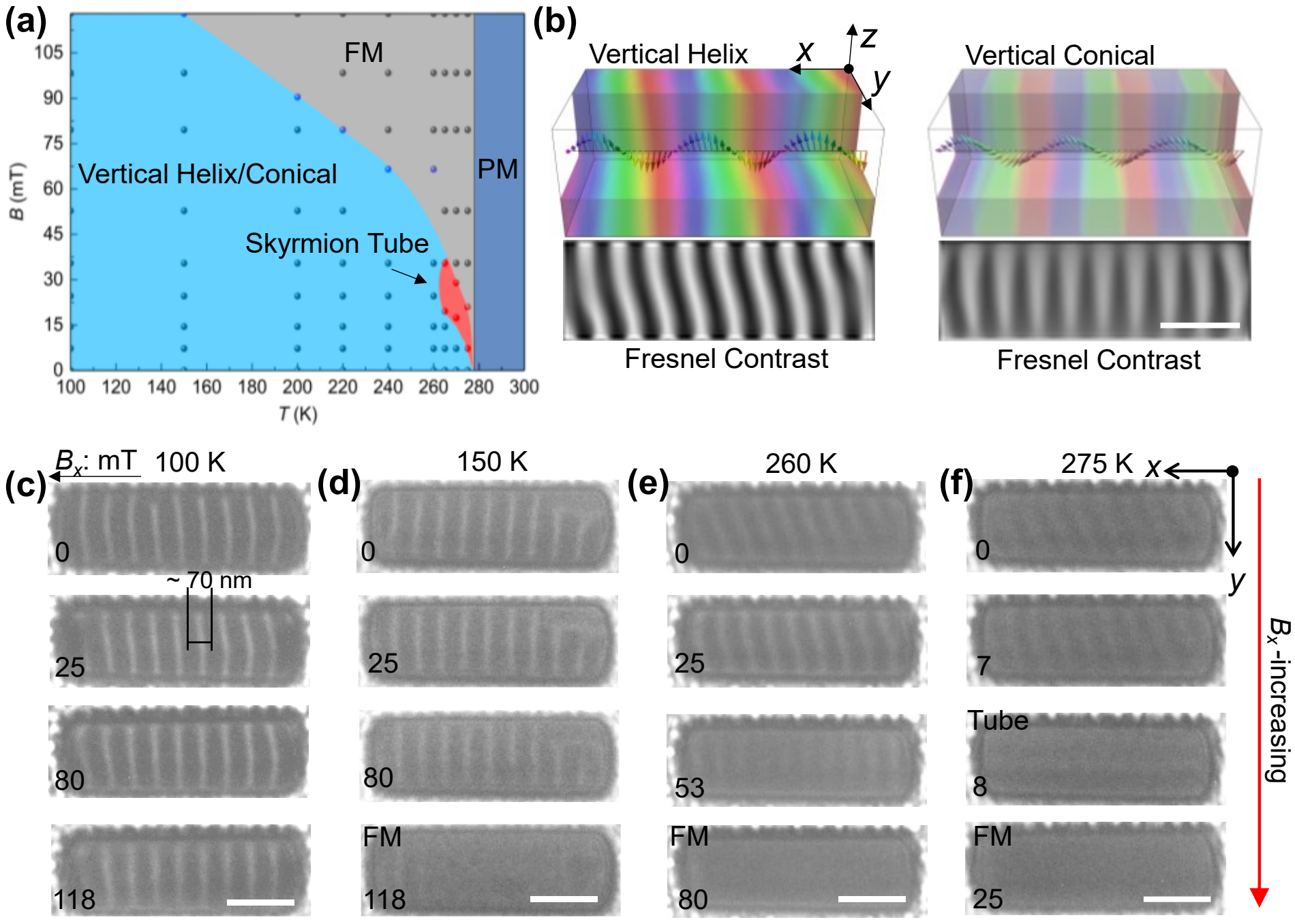

**Figure 2.** Magnetic evolutions of vertical helix ground state in a FeGe cuboid with $l$ = 700 nm. (a) Experimental magnetic thermal equilibrium phase diagram of the FeGe cuboid. FM refers to the ferromagnetic state and PM refers to the paramagnetic state. Dots represent full experimental field intervals when recording magnetic structures. (b) Three-dimensional schematic diagrams of representative simulated magnetic states including vertical helix, vertical conical, and their corresponding Fresnel contrasts. (c)–(f) Magnetic evolutions under an in-plane magnetic field at 100, 150, 260, and 275 K. Scale bars, 200 nm.

At zero field, helix domains always exist as magnetic ground states with a $q$-vector oriented almost along the length orientation ($x$-axis) of the FeGe cuboid below Curie temperature. We refer to this as the vertical helix. At first, the $q$-vector of the helix with a period of approximately 70 nm (Figure 2c) may deviate slightly from the $x$-axis (Figure 2c-f). Once in-plane fields along the $x$-axis $B_x$ are applied, the $q$-vectors of magnetic configurations gradually align with the $x$-axis (Figure 2c–f), which is well reproduced in micromagnetic simulations based on measured magnetic parameters of FeGe (Figure S2). Moreover, as the in-plane field $B_x$ is applied, the zero-field helix transforms to conical domains, with *q*-vectors coinciding with the field orientation, *i.e.*, the *x*-axis, leading to the reorientation of *q*-vectors. At temperatures below 260 K, in the $B_x$-increasing process, the vertical helix transforms into the conical, and finally into the ferromagnet (FM). Note that the conical-to-FM transformation occurs first in the central regions and then spreads to the edge regions. Thus, at a high field, FM can be observed in the interior of cuboids while conical domains can still exist near two edges (Figure 2d). Such an evolution driven by in-plane fields is also consistent with our zero-temperature simulations (Figure S2). In our zero-temperature simulations and experiments at temperatures below 260 K, no skyrmion tubes can be created from the vertical helix. The transformations from the vertical helix to the conical state can be considered second-order magnetic phase changes, characterized by continuous transitions with magnetizations gradually titled toward the *x*-axis (Figure 2b).

In a narrow temperature range of 265–275 K near $T_c$, the vertical helix transforms horizontal skyrmion tubes (Figure 2f), associated with a sudden reorientation of the *q*-vectors from the *x*-axis to the *y*-axis, during the application of in-plane fields. Although the horizontal skyrmion tubes and horizontal helix exhibit identical Fresnel contrasts, they can be distinguished based on the intensity of these contrasts (Figure 3a, c). The magnetizations along the depth orientation (*z*-axis) reverse for skyrmion tubes, leading to a significant cancellation of magnetizations within the *xy* plane (Figure S3). Consequently, the Fresnel contrasts of horizontal skyrmion tubes are considerably weaker than those of horizontal helix (Figure 3b, d and Figure S3). We distinguish skyrmion tubes based on the sudden weakened Fresnel contrasts ($B_x$ = 8 mT for Figure 2f). It is noteworthy that the Fresnel contrast is also influenced by temperature and magnetic field. Higher temperatures lead to a decrease in saturated magnetization, resulting in a weaker Fresnel contrast. Likewise, increasing the magnetic field gradually makes the magnetization more uniform, which causes a reduction in the Fresnel contrast. Therefore, to ensure accurate differentiation of skyrmion tubes under similar conditions, we prefer to analyze cases where coexisting skyrmion tube and horizontal helix are present within the same cuboid, both tubes and helix are subject to the same magnetic field, and when the field undergoes minimal changes resulting in a transformation from helix to tubes (Figure S4). When further increasing the field, the tubes turn to FM without an intermediate conical phase (Figure 2f). The sudden transformations from the vertical helix to the tube, and finally FM are first-order phase changes.

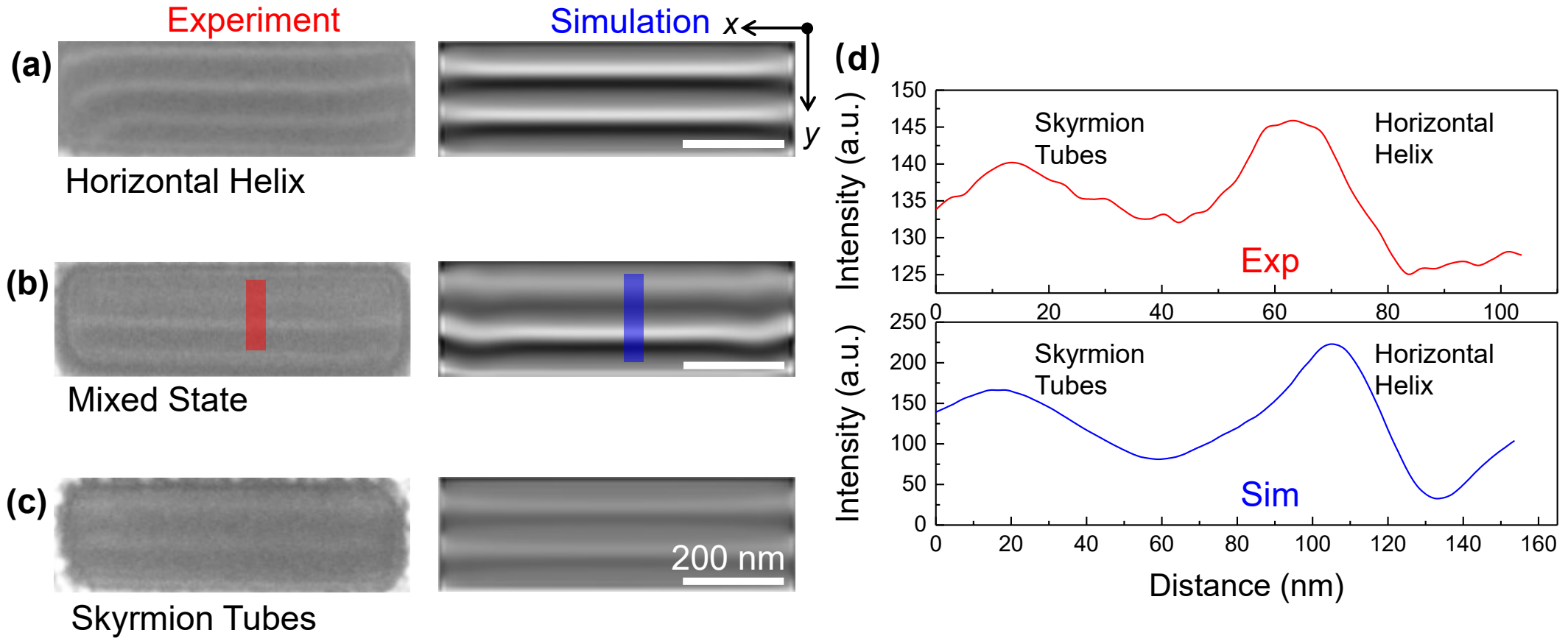

**Figure 3.** Characteristic magnetic states recorded in experiments and simulations. (a) Experimental Fresnel contrast of horizontal helix in comparison with simulated results. (b) Experimental Fresnel contrast of mixed state (a skyrmion tube and a horizontal helix) in comparison with simulated results. (c) Experimental Fresnel contrast of horizontal helix in comparison with simulated results. The defocus distance is -0.35 mm. The scale bar is 200 nm. (d) Distance dependence of Fresnel contrast intensity for the skyrmion tube in (b).

We obtain the complete experimental magnetic diagram of the FeGe cuboid as a function of temperature and field, starting from the magnetic ground vertical helix at zero magnetic fields, as shown in Figure 2a. we incrementally increased the magnetic field in steps up to 118 mT at various temperatures, recording the critical transformation fields between different magnetic states. The phase diagrams is then plotted based on these critical transformation fields.

Representative three-dimensional magnetic configurations of the vertical helix and vertical conical are shown in Figure 2b. For FeGe cuboids with other lengths, the stabilization of skyrmion tubes does not exhibit significant differences (Figure S5 and S6).

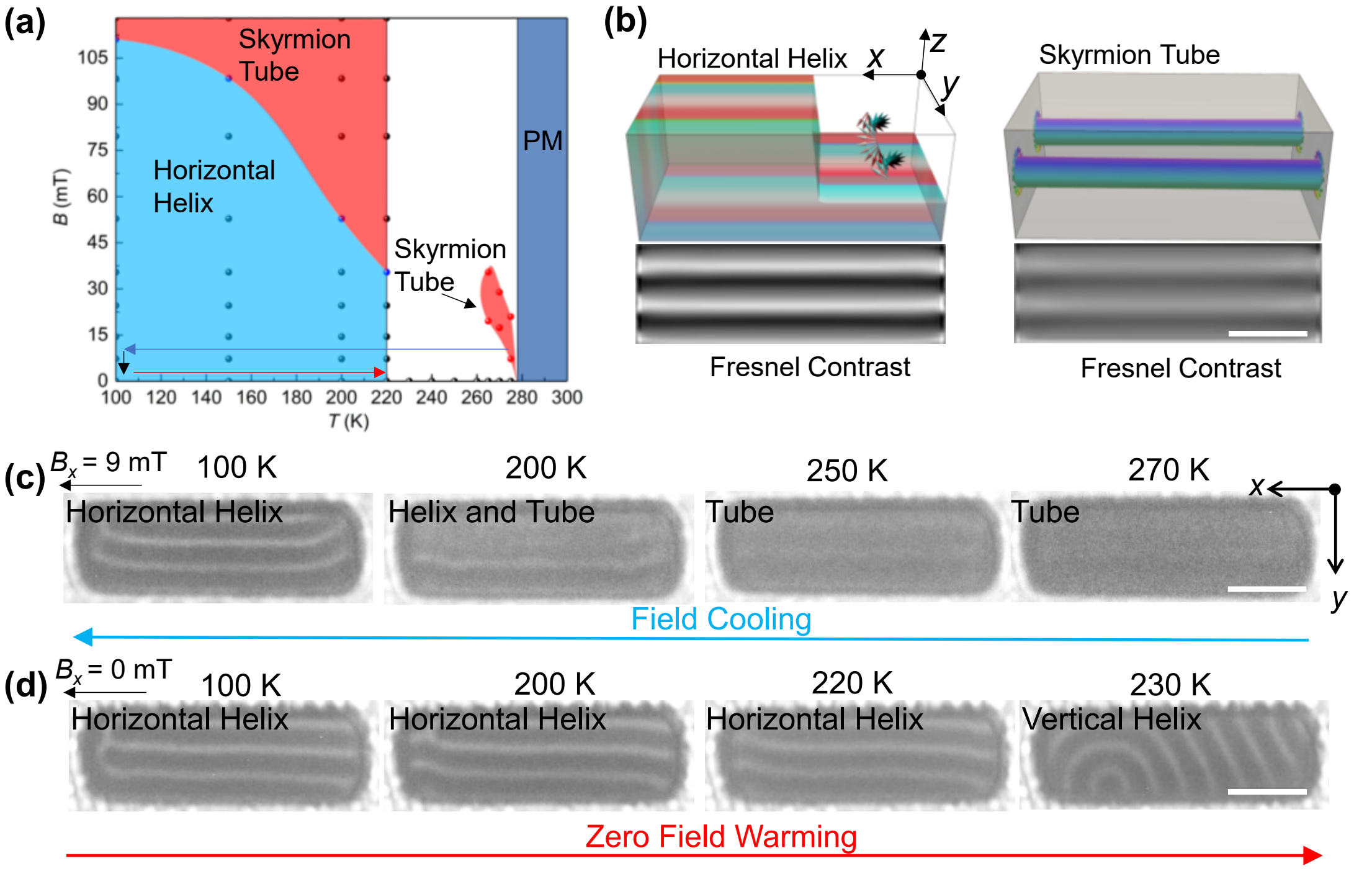

**Figure 4.** Magnetic evolutions of horizontal helix ground state in a 700 nm FeGe cuboid. (a) Magnetic phase diagram of metastable skyrmion tubes. (b) Three-dimensional schematic diagrams of representative simulated magnetic states including horizontal helix, skyrmion tube, and their corresponding Fresnel contrasts. (c) Typical Lorentz images at different temperatures (100, 200, 250, and 270 K) during the field cooling process. Scale bar, 200 nm. (d) Typical Lorentz images at different temperatures (100, 200, 220, and 230 K) during the zero-field warming process. Scale bar, 200 nm.

Despite that the vertical helix phase always remains stable at zero fields in the FeGe cuboid, we can achieve a horizontal helix with a *q*-vector along the *y*-axis through a field-cooling process, as shown in Figure 4. We first apply a small in-plane field $B_x$ = 9 mT to stabilize two skyrmion tubes at 270 K, then cool the FeGe cuboid to 100 K. During the field-cooling process, the two skyrmion tubes in the FeGe cuboid remain stable for temperatures above 200 K (Figure 4c). At $T$ = 200 K, one of the tubes transforms to the helix with strongly enhanced magnetic contrast (Figure 4c). Upon further reduction in temperature to 100 K, only the horizontal helix phase is observed. It is worth noting that the horizontal helix also exists as a metastable phase when the in-plane field decreases to zero. Three-dimensional structures of horizontal helix and skyrmion tubes are displayed in Figure 4b and the two magnetic textures have the same $q$-vector, which is parallel to the $y$-axis. Additionally, we investigate the stability of the horizontal helix as a function of temperature at zero field. During the field-free warming process, the horizontal helix remains stable but turns to the vertical helix for temperatures exceeding 220 K (Figure 4d).

Subsequently, we examine the magnetic evolution driven by in-plane fields starting from the horizontal helix at $T \leq 220$ K (Figure S7). Our findings indicate that the Fresnel contrasts of tubes are considerably weaker compared to those of helix. As the in-plane magnetic fields increase, the Fresnel contrasts undergo a sudden decreased intensity in the same defocused condition, suggesting the transformation from horizontal helix to tubes. To further investigate this phenomenon, we conducted simulations of the magnetic evolution from the horizontal helix at zero temperature (Figure S8a). The variations in Fresnel contrast during the transformation from horizontal helix to tubes (Figure S8b) are excellently consistent with experiments. In our experiments, two skyrmion tubes can be always obtained at high fields from the horizontal helix in the FeGe cuboid at $T \leq 220$ K. The threshold field for the helix-to-tube transformation decreases as the temperature increases. Furthermore, our experiments reveal that the skyrmion tubes exhibit greater stability against in-plane fields compared to vertical conical domains. For example, the vertical conical domains transform into an FM state at 118 mT at 200 K (Figure S9). Whereas the two skyrmion tubes remain stable under the same conditions (Figure S7c).

We have examined the stability of skyrmion tubes in the FeGe cuboid as a function of temperature and field, as shown in Figure 4a. At high temperatures ranging from 265 to 275 K, skyrmion tubes are spontaneous magnetic ground states within certain field ranges and the simulated three-dimensional structure of the tubes is demonstrated in Figure S10. At low temperatures from 100 to 220 K, we show the achievement of skyrmion tubes through the transition of the metastable horizontal helix at zero fields. However, the stability of skyrmion tubes within the temperature range from 220 to 260 K is not displayed due to the absence of the horizontal helix at zero fields (Figure 4d). Figure 4b shows the simulated magnetization of the middle layer and the corresponding Fresnel contrast of both the horizontal helix and skyrmion tubes. Thus, we have demonstrated the Fresnel contrast difference between the skyrmion and the horizontal helix under different temperatures (Fig. S7), magnetic field ranges (Fig. S7), and samples (Fig. S6).

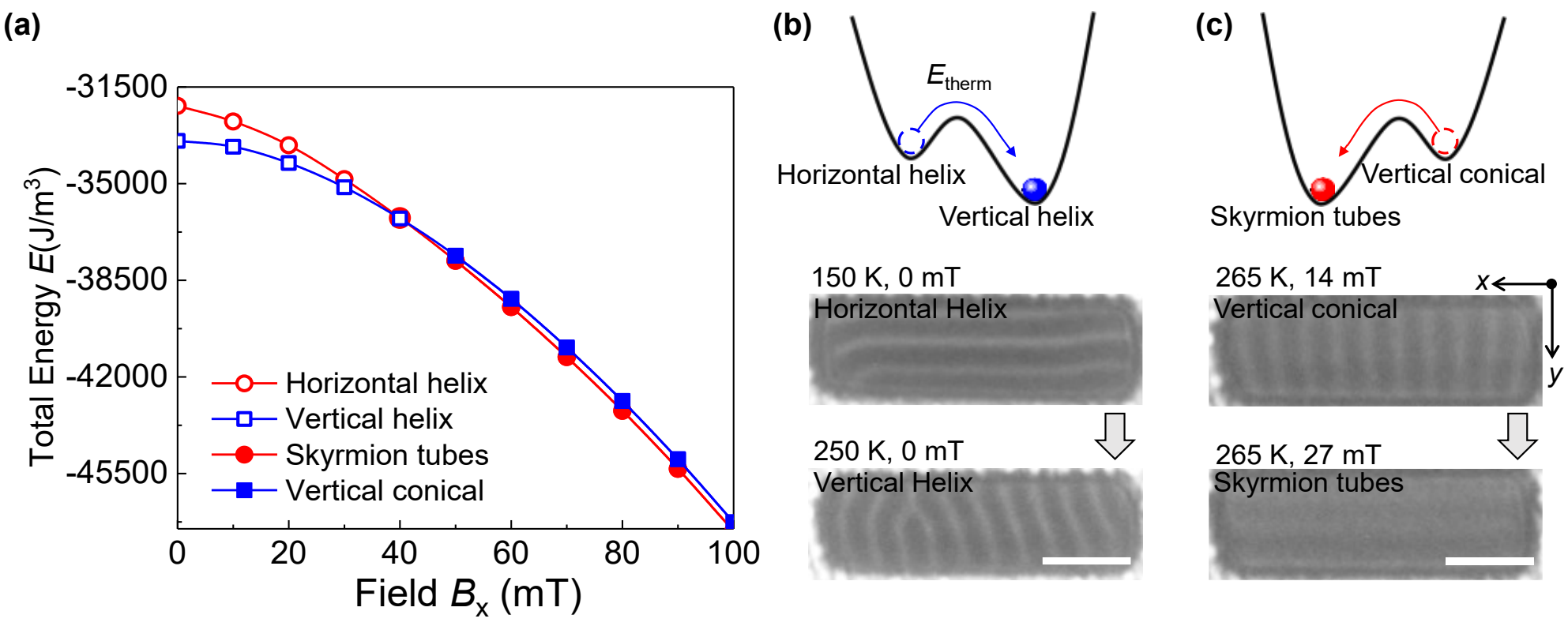

**Figure 5.** The mechanism of the transformation between magnetic states. (a) Simulated total energy $E$ for different magnetic states as a function of magnetic field $B_x$. (b) Transformation from a horizontal helix to a vertical helix with an increasing temperature at $B_x$ = 0 mT. (c) Transformation from vertical conical to skyrmion tubes at 265 K. Scale bar, 200 nm.

To investigate the stability of spin textures in the nanostructured FeGe cuboid, we analyze the energy profiles during the magnetic evolution between the magnetic states with the $q$-vector along the $x$-axis or $y$-axis, as shown in Figure 5a. At zero fields, the horizontal helix (with the $q$-vector along the $y$-axis) exhibits significantly higher total energy compared to the vertical helix (with the $q$-vector along the $x$-axis). When the temperature approaches $T_c$, the enhanced thermal fluctuation energy facilitates the transition from a high-energy horizontal helix to a low-energy vertical helix (Figure 5b), which explains the instability of the horizontal helix at high temperatures (Figure 4d). When applying in-plane fields, the horizontal helix transforms into skyrmion tubes while the vertical helix converts into vertical conical domains. The energy profile reveals that the skyrmion tubes possess lower energy than the vertical conical domains (Figure 5c). Thus, the high thermal fluctuation energy near $T_c$ assists the transformation from high-energy conical domains to low-energy skyrmion tubes at high magnetic fields (Figure 2f and Figure S5).

At low temperatures, thermal fluctuation energy is not dominant, the high-field magnetic phases are directly correlated with initial magnetic states at zero fields with no reorientation of $q$-vectors. Specifically, the vertical helix converts to vertical conical domains, while the horizontal helix transforms into skyrmion tubes. However, at high temperatures near $T_c$, the strong thermal fluctuation energy contributes to the instability of metastable phases, leading to the absence of hysteresis. Therefore, for the high-temperature range of 265−275 K, the spin textures have a $q$-vector along the $x$-axis (the vertical helix) at low fields, and spin textures have a $q$-vector along the $y$-axis (the skyrmion tubes) and can be mutually transformed reversibly at high fields.

In summary, we investigated the magnetic evolutions of skyrmion tubes in confined FeGe cuboids under an *in-situ* in-plane magnetic field. Our results show that thermally stable skyrmion tubes only exist within a narrow field-temperature window near $T_c$. By using a field cooling method, a horizontal helix with a $q$-vector perpendicular to the length orientation ($x$-axis) of the cuboid can be achieved. Subsequently, through the application of an appropriate magnetic field $B_x$, the horizontal helix transforms into skyrmion tubes. These magnetic state transformations are primarily driven by competitive energies and thermal fluctuations. Our findings contribute to a better understanding of three-dimensional topological magnetic structures in confined geometries and have potential implications for the use of skyrmion tubes in advanced devices.

## Methods

**Sample preparation.** FeGe single crystals with B20-structure were grown by chemical vapor transport method with stoichiometric iron (>99.9%), germanium (>99.9%), and transport agent $I_2$. Under a temperature gradient from 560 °C to 500 °C, FeGe crystallized with the $P2_13$ space group, forming pyramidal-shaped crystals. The nanostructured FeGe cuboids with different lengths were fabricated via a standard lift-out method from the bulk material by using a focused ion beam instrument (Helios Nanolab 600i, FEI).

**TEM Measurements.** Fresnel magnetic images were recorded by a Lorentz TEM instrument (JEM-2100F), of which the objective lens is specially designed to decrease the field at the sample position. The in-plane magnetic field was applied by a self-made in-situ specimen holder, which can offer the largest magnetic field at about 118 mT.

**Micromagnetic Simulations.** Micromagnetic Simulations were performed by using a GPU-accelerated micromagnetic simulation program, Mumax3.[54] The exchange energy, Dzyaloshinskii-Moriya interaction (DMI) energy, Zeeman energy, and demagnetization energy were considered in the simulations. Magnetic parameters were set based on the FeGe material with exchange stiffness $A_{ex}$ = 3.25 pJ/m, and saturation magnetization $M_s$ = 384 kA/m.[28] The DMI interaction $D_{dmi} = 4\pi A_{ex}/L_D$ = 5.834 mJ $m^{-2}$ was obtained from the zero-field spin spiral period $L_D$ = 70 nm.[28] The cell size was set to 4 × 4 × 4 $nm^3$. The equilibrium spin configurations were obtained by using the conjugate-gradient method.

**Associated content**

The authors declare no competing financial interest.

**Supporting Information**

The Supporting Information is available free of charge on the ACS Publications website at http://pubs.acs.org.

**Figures S1 to S3.** Magnetic phase diagram in increased and decreased field processes. Simulated field-driven magnetic evolution in the $Fe_3Sn_2$ cuboid. Pulse duration and magnetic field dependence of threshold current densities required for the creation $j_{c1}$ and deletion $j_{c2}$. **Video S1 to S3.** Simulated STT-driven FM-to-skyrmion transformation. Simulated skyrmion-to-FM transformation. Writing and deleting single skyrmions using single 80-ns pulsed currents with densities of 8.6 and 12 × $10^{10}$ A $m^{-2}$, respectively.

**Author information**

Corresponding author

*(J.T.) Email: jintang@ahu.edu.cn; *(W.X.) Email: xiawxing@nimte.ac.cn; *(H.D.)

Email: duhf@hmfl.ac.cn

**Authors**

**Jin Tang-** https://orcid.org/0000-0001-7680-8231

**Acknowledgments**

This work was supported by the National Key R&D Program of China, Grant No. 2022YFA1403603; the Natural Science Foundation of China, Grants No. 12174396, 12104123, and 12241406; the National Natural Science Funds for Distinguished Young Scholar, Grants No. 52325105; Anhui Provincial Natural Science Foundation, Grant No. 2308085Y32; the China Postdoctoral Science Foundation, Grant No. 2023M743543; Natural Science Project of Colleges and Universities in Anhui Province, Grant No. 2022AH030011; the Strategic Priority Research Program of Chinese Academy of Sciences, Grant No. XDB33030100; CAS Project for Young Scientists in Basic Research, Grant No. YSBR-084; and Ningbo Key Scientific and Technological Project, Grants No. 2021000215 and 2023Z099.

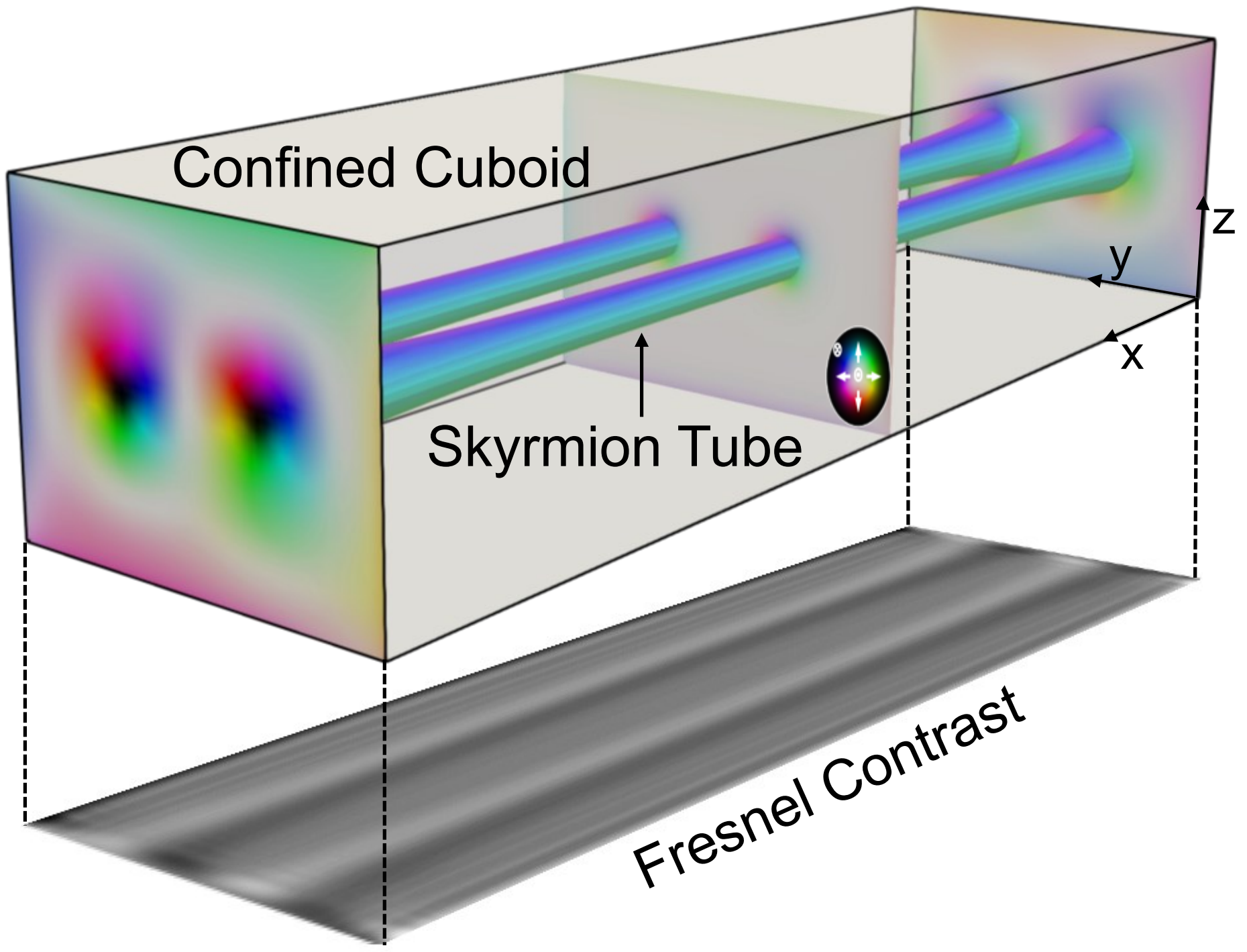

**For table of Contents Only**